\documentclass[aip,pop,reprint]{revtex4-1}

\usepackage{graphicx}
\usepackage{amssymb}
\usepackage{amsmath}
\usepackage{hyperref}
\usepackage{color}

\newcommand{\omegb}{\omega_{\nabla B}}
\newcommand{\omecv}{\omega_{k}}
\newcommand{\fgb}{f_{\nabla B}}
\newcommand{\fcv}{f_{k}}
\newcommand{\aky}{k_y\rho_i}

\begin{document}

\title{On the linear stability of collisionless microtearing modes}
\author{I. Predebon, F. Sattin}
\affiliation{Consorzio RFX, Associazione EURATOM-ENEA sulla Fusione, Padova, Italy}
\begin{abstract}
Microtearing modes are an important drive of turbulent heat transport in present-day fusion plasmas. We investigate their linear stability under very-low collisionality regimes, expected for the next generations of devices, using gyrokinetic and drift-kinetic approaches. At odds with current opinion, we show that collisionless microtearing instabilities may occur in certain experimental conditions, particularly relevant for such devices as reversed field pinches and spherical tokamaks.
\end{abstract}
\pacs{52.35.Qz,52.65.Tt}
\date{\today}
\maketitle


Microtearing modes (MTMs) are short-wavelength electromagnetic instabilities driven by electron temperature gradients, investigated since the 1970s for their relevance to heat transport in fusion plasmas~\cite{hazeltine}. Interest toward them has strongly revived in the last decade thanks to the widespread adoption of sophisticated gyrokinetic codes running on powerful computers, allowing the study of MTMs in realistic geometries, including tokamaks, spherical tokamaks, and reversed field pinches (RFPs)~\cite{mast,nstx,rfx}. Recently the first nonlinear simulations of MTMs~\cite{nonlin1,nonlin2} were able to show that these modes can be responsible for a large fraction of turbulent heat transport in tokamaks. Apart from the plasma core region, MTMs are thought to be relevant also in the edge as players regulating the heat transport, and possibly the pedestal evolution between ELMs~\cite{hmode}.

Even more important is to assess how MTMs can impact future plasma conditions, in the quest towards reactor-grade scenarios. To a large extent, this amounts to investigating hotter and thus less collisional plasmas. The reference works addressing the role of collisionality $\nu$ are Drake {\it et al}~\cite{drake}, and Gladd {\it et al}~\cite{gladd}. Their conclusion is that -- all the other parameters fixed -- the growth rate peaks at a finite value of $\nu$ and decreases down to negative values as $\nu \to 0$.

This basic picture held until nowadays, just marginally modified by subsequent studies by Catto and Rosenbluth~\cite{catto} and Garbet {\it et al}~\cite{garbet}, and is apparently supported by gyrokinetic simulations~\cite{mast}. Very recently, on the other hand, several papers have been accumulating a consistent amount of evidence in favor of the opposite view: MTMs may actually be destabilized even under vanishingly small-$\nu$ conditions~\cite{vac2,vac3,vac4,vac5,vac6}. However this disagreement is a consequence of the fact that a {\it generic} collisionless regime does actually not exist: $\nu$ is not the only important non-ideal mechanism; its role may be played, for example, by particle inertia. These mechanisms were either unaccounted for in earlier works, or the plasma conditions addressed were not suitable for their development.

The purpose of this Letter is to investigate the occurrence of collisionless MTMs, considering previously unaccounted destabilizing mechanisms/regimes. They include the magnetic drifts in the kinetic equations, and the mutual role of particle density and magnetic shear profiles. At first we will make an assessment on MTM linear stability using the gyrokinetic code GS2~\cite{gs2} in the experimental scenarios more prone to MTM turbulence, in particular for the reversed field pinch geometry. Afterwards we will validate such results with a slab drift-kinetic model, and justify our conclusions as relevant in particular for high magnetic drift configurations. Besides yielding an independent cross-check of the results, the drift-kinetic model, although less accurate quantitatively, provides a deeper insight into the physical mechanisms behind the mode destabilization.


{\it Gyrokinetic approach.} The gyrokinetic equation is solved in a flux-tube domain by means of the electromagnetic code GS2 in RFP geometry~\cite{gs2rfp}. Unlike the tokamak configuration, the RFP is characterized by a fast winding of the magnetic field in the poloidal direction, which causes the safety factor profile $q$ to be $\ll 1$, and to vanish in the very edge of the plasma. This reflects on a different ordering of the drift frequency $\omega_d$ in the gyrokinetic equation. Given $B$ the magnetic field strength (normalized to the average field on the flux surface), with $B_\theta$ and $B_\phi$ the poloidal and toroidal component, the gradient $\omegb$ and curvature $\omecv$ terms in $\omega_d$ oscillate around $-B'$ (hereafter $\cdot'\equiv d\cdot/dr$) and $B_\theta^2/r$ respectively as a function of the poloidal (ballooning) angle $\theta$, see Ref.~\onlinecite{gs2rfp}. Thus the average normalized $\omegb$ and $\omecv$ are not of the same order as the inverse aspect ratio $a/R$, like in the tokamak, but are typically larger than 1.

For our study we start from a reference internal-transport-barrier scenario of RFX-mod, already investigated in Ref.~\onlinecite{rfx}: experimental values of safety factor $q\simeq 1/10$, magnetic shear $\hat{s}=rq'/q\simeq -0.6$, normalized logarithmic density gradient $a/L_n=-an'/n=0.2$, with $a$ torus minor radius; for the ion/electron temperature gradients we assume $a/L_{T_i}=0$ (in order to limit the number of potential instabilities, this quantity is actually unknown in RFX-mod) and $a/L_{T_e}\simeq 4$, respectively. The plasma $\beta$, the parameter ruling the electromagnetic effects in the gyrokinetic equation, is artificially increased with respect to the experimental value ($\beta_e$ approximately 0.7\%), so as to work well above the stability threshold. Particle collisions are included by means of a classical Lorentz operator, fitted to describe pitch angle scattering and trapped-passing particle interactions; the momentum exchange between species (electrons and ions), together with the energy scattering and slowing down processes, is not considered in GS2. For the numerical simulations we include fluctuations in the parallel vector potential $A_\parallel$. The electrostatic potential $\phi$ is usually retained, although it may be artificially switched off in some cases, so as to assess its impact upon final results. Throughout this section, $\nu$ is normalized to its experimental value at the plasma conditions considered, $\nu_\mathrm{exp}\simeq 0.4\,v_{\mathrm{th},i}/a$, with $v_{\mathrm{th},i}$ ion thermal speed. Concerning the resolution of our calculations, the longitudinal grid must be large enough to resolve the elongated structure of the MTM eigenfunctions, $|\theta|\leq 60\pi$ with 32 grid points per $2\pi$-period; for the velocity space we typically set 12 untrapped pitch angles and 16 values of energy, having performed convergence tests with increased number of grid points on a subset of cases at the lowest collisionalities.

\begin{figure}
\includegraphics{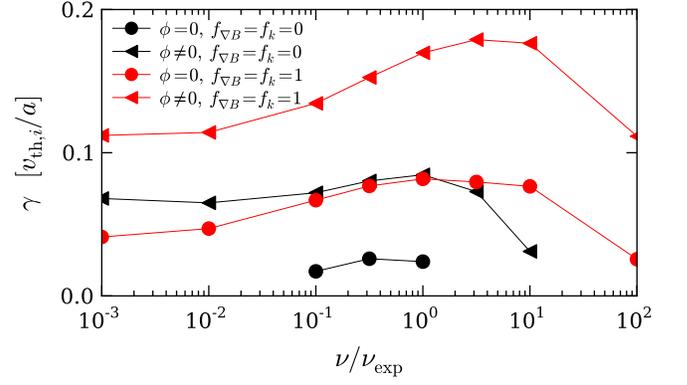}
\caption{Growth rate as a function of collisionality, with and without electrostatic fluctuations, with ($\fgb=\fcv=1$) and without ($\fgb=\fcv=0$) $\omegb$ and $\omecv$ drifts. Relevant parameters: $a/L_{T_e}=4$, $a/L_n=0$, $\hat{s}=-0.65$, $q=0.12$, $\beta_e=0.05$, $\aky=0.1$.}
\label{fig:gammanu}
\end{figure}

For all such parameters the fastest growing instability turns out to be of MTM type, in the range $\aky\lesssim 1$. Sometimes, another branch with opposite parity appears at higher wavenumbers, due to the passing electron response, similar to that described in Ref.~\onlinecite{hall}. The microtearing nature of the modes is clearly evident by looking at the mode structure (odd symmetric and very elongated in $\phi$, even and localized in $A_\parallel$), at the sign of the real frequency (corresponding to a propagation in the electron diamagnetic direction), and at the trend of the growth rate as a function of electron temperature gradient and plasma $\beta$. The growth rate definitely does not vanish when decreasing collisionality, as shown in Fig.~\ref{fig:gammanu}.

The most relevant difference between the slab model used in Refs.~\onlinecite{drake,gladd} and the RFP geometry is the complexity of the $\omega_d$ drift in the latter. In the following we investigate its role in destabilizing MTMs, having defined $\fgb,\fcv\in[0,1]$ as parameters artificially weighting the fraction of $\omegb$ and $\omecv$ respectively, keeping $\fgb=\fcv$ for the moment. In Fig.~\ref{fig:gammanu} we show the growth rate of the fastest growing MTM for a flat-density ($a/L_n=0$) high-beta plasma. While the curves have a maximum at a given $\nu$, the growth rate is non-vanishing also for $\nu \to 0$, except for the case without $\phi$ fluctuations and no drifts. In agreement with Ref.~\onlinecite{mast}, retaining the electrostatic potential is always destabilizing. Large growth rates for $\fgb=\fcv=1$ generally correspond to upshifted $\gamma(k_y)$ spectra, peaked for $\aky>1/2$; otherwise, the peak occurs around $\aky=0.1-0.3$.

\begin{figure}
\includegraphics{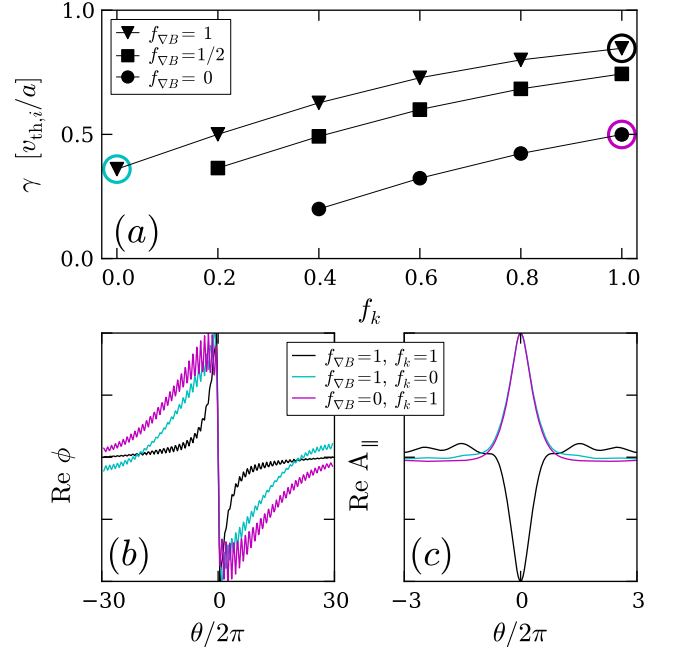}
\caption{Growth rate as a function of fractions of $\omegb$ and $\omecv$ (a) with some representative eigenfunctions of electrostatic potential (b) and parallel vector potential (c), for $\nu=10^{-6}$. Relevant parameters: $a/L_{T_e}=4$, $a/L_n=0.2$, $\hat{s}=-0.65$, $q=0.12$, $\beta_e=0.08$, $\aky=0.5$.}
\label{fig:gammadrift}
\end{figure}

\begin{figure}
\includegraphics{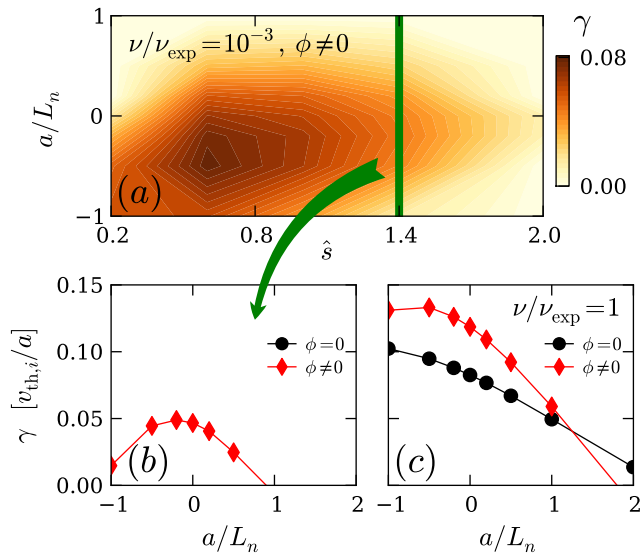}
\caption{Growth rate vs. magnetic shear and density gradient for $\nu=10^{-3}$ (a.u.), including electrostatic fluctuations (a). The section $\hat{s}=1.4$ is shown in (b). The same section for $\nu= 1$ is in (c). Relevant parameters: $a/L_{T_e}=4$, $q=0.12$, $\beta_e=0.05$, $\fgb=0=\fcv$, $\aky=0.2$.}
\label{fig:gammashear}
\end{figure}

It is instructive to quantify how the single drifts separately modify the destabilization of MTMs, Fig.~\ref{fig:gammadrift}. For this study we assume $\phi\ne 0$, a higher wavenumber and plasma beta, $(\aky=0.5, \beta_e=0.08$), and a very low collisionality, $\nu=10^{-6}$. The contribution of the two drifts turns out to be comparable; they separately provide a monotonic increase of the growth rate. For $\fgb=0=\fcv$, at this value of $\aky$, MTMs are stable. Note that this is due to the mentioned downshift of the cutoff of the spectrum $\gamma(k_y)$ as $\fgb,\fcv\to 0$.

Up to this point we may claim that devices featuring high drifts are particularly prone to MTMs. Besides drifts, however, we identify another destabilizing mechanism, which arises due to the mutual balance between magnetic shear, temperature and density gradient. In Fig.~\ref{fig:gammashear} we show the growth rate as a function of magnetic shear $\hat{s}$ and logarithmic density gradient $a/L_n$ for a low collisionality plasma, $\nu=10^{-3}$, and with $\fgb=\fcv=0$. Furthermore, the trapped particle fraction is set to 0, so as to exclude further destabilizing contributions~\cite{catto}. We find the existence of a maximum of $\gamma$ in the $a/L_n<0$ half-plane: positive density gradients are increasingly stabilizing MTMs for each value of the magnetic shear, Fig.~\ref{fig:gammashear}-a/b, whereas slightly negative gradients provide destabilization. An analogous behaviour of $\gamma(a/L_n)$ has been recently discussed in Ref.~\onlinecite{gutten} for (collisional) spherical tokamak plasmas. On the other hand, for a fixed density gradient the growth rate has a maximum in $\hat{s}$. The appearance of this maximum is not unexpected, since a similar dependence on the parameter $L_n/L_s$, with $L_s=qR/\hat{s}$ magnetic shear length, was already encountered in Ref.~\onlinecite{gladd} and again in Ref.~\onlinecite{gutten}; the difference here is that we find a positive $\gamma$ also for $\nu\ll 1$. Notice that, at this level of collisionality, the only way to destabilize MTMs is to include $\phi$ fluctuations. The same study performed keeping $\phi=0$ identically provides MTM stability for every couple $(a/L_n,\hat{s})$. Increasing collisionality further destabilizes the mode, as expected: in Fig.~\ref{fig:gammashear}-c the function $\gamma(a/L_n)$ is shown for $\nu = 1$. While the roles of the density gradient and of the magnetic shear are confirmed, we notice that the maximum in $a/L_n$ is moving to much lower (unphysical) values, and that, for this collisionality, the modes can be destabilized even with $\phi=0$. These results do not contradict Refs.~\onlinecite{drake,gladd}, since we are in a different regime: those papers addressed the case of a plasma featuring strong temperature gradients and more moderate but still notable density gradients, $\eta=L_n/L_{T_e}>1$, $\xi=L_n/L_s\ll 1$, which is a fairly reasonable picture of a device centrally heated and fuelled. On the other hand, while a reactor-grade device must feature relevant temperature gradients, density profiles could be rather flat in the absence of relevant central particle sources, as happens in devices without neutral beams.


{\it Drift-kinetic approach.} We solve the Amp\`ere equation and the quasi-neutrality equation in slab geometry, as extensively explained in Refs.~\onlinecite{drake,gladd}:
\begin{eqnarray}
\label{eq:dka}
&& A''_\parallel - k_y^2{A_\parallel} = \psi\sigma \left(\omega {A_\parallel} - x\,\varphi \right) \\
\label{eq:dkp}
&& (1+ \omega) \left(\varphi'' - k_y^2\varphi \right) = \chi x \psi\sigma \left( \omega {A_\parallel} - x\,\varphi \right)
\end{eqnarray}
where $\chi = \xi^2 (c_a/c)^2$, $\psi = (8/3\sqrt\pi)(\omega_\mathrm{pe}\rho_i/c)^2$, $\varphi = \phi\,\xi\,(c/u_i)$, and $\cdot'=d\cdot/dx$. The parallel conductivity $\sigma$ is defined in the original papers \cite{drake,gladd}. Lengths are normalized to the ion thermal Larmor radius $\rho_i$ and frequencies to the electron diamagnetic frequency $\omega_*$; $x$ is a radial-like coordinate orthogonal to the equilibrium magnetic field $B_0 \hat z$, $x=0$ being the resonance position of the mode; $k_y$ and $k_\parallel = k_y x/L_s$ are respectively the component of the wavenumber of the mode orthogonal and parallel to the magnetic field; $c, c_a, u_i$ the light, Alfv\'en and ion thermal speed, $\omega_{pe}$ the plasma frequency. Equations (\ref{eq:dka}) and (\ref{eq:dkp}) are to be solved with the boundary conditions satisfying the symmetry of the MTM: $A'(0)=0$, $\phi(0)=0$, $A'(\infty)/A(\infty)=\phi'(\infty)/\phi(\infty)=-k_y$. This yields four real-valued eigenvalue problems for the quantities $\omega = \omega_r + i \gamma$ (the eigenfrequency of the mode) and Re$(\phi'(0))$, Im$(\phi'(0))$, that are determined via Newton-Raphson method. Our solver of Eqns. (\ref{eq:dka},\ref{eq:dkp}) has been first benchmarked against a selected set of results picked up from Refs.~\onlinecite{drake,gladd}.

\begin{figure}
\includegraphics[width=7cm]{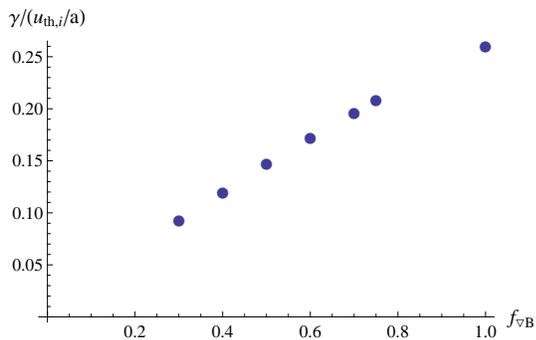}
\caption{Growth rate {\it versus} $\nabla B$ drift (for the definition of $\fgb$ see the main text). Other parameters are: $\nu = 0, k_y = 0.05, \beta = 10^{-2}, \eta = 1 , \xi = 0.1$ }
\label{fig:dk5}
\end{figure}

According to the previous section, let us first address the issue of the drifts. The slab geometry by definition cannot accommodate curvature drifts, but $\nabla B$ ones can be included heuristically: $\vec{u}_{\nabla B} \approx - (m_e u_\perp^2 /2e B_0) (x/L_s^2) \hat{y} + O(x^2) \hat{z}$. At this stage, $u_{\nabla B}$ is defined apart from numerical factors of order unity; it can be conveniently written in dimensionless form as $u_{\nabla B}=\fgb x L_n/L_s$, where the factor $\fgb$ has been defined above, and is used to modulate the effect of the $\nabla B$ drift. The collisionless linearized drift-kinetic equation becomes
\begin{eqnarray}
\label{eq:dke}
&& \left[ i\omega - ik_{\parallel} u_{\parallel}  - ik_y  u_{\nabla B} \right] f = \nonumber \\
&& = e E_\parallel u_\parallel f_0/T - (i k_y c/B_0) (\phi - A_{\parallel} u_{\parallel} /c) f'_0 \,,
\end{eqnarray}
where $f_0$ is the equilibrium electron distribution function, and $f$ its perturbed part. The first moment over the velocity yields the perturbed current used in the Amp\`ere and quasi-neutrality equations. Indeed, by increasing $\fgb$ the drift term becomes more and more important: an approximately linear increase of $\gamma$ with $\fgb$ is featured, as is shown in Fig.~\ref{fig:dk5}. The physical picture is that non-ambipolar drifts, such as $\nabla B$ and curvature, provide the current across the magnetic field needed for reconnection.

\begin{figure}
\includegraphics[width=7cm]{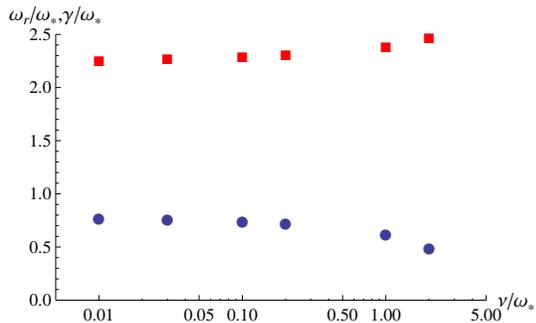}
\caption{Real (red squares) and imaginary (blue dots) part of the eigenfrequency {\it versus} collision frequency. Other parameters are $k_y = 0.05, \beta = 10^{-2}, \eta = \xi = 6$.}
\label{fig:dk1}
\end{figure}

\begin{figure}
\includegraphics[width=7cm]{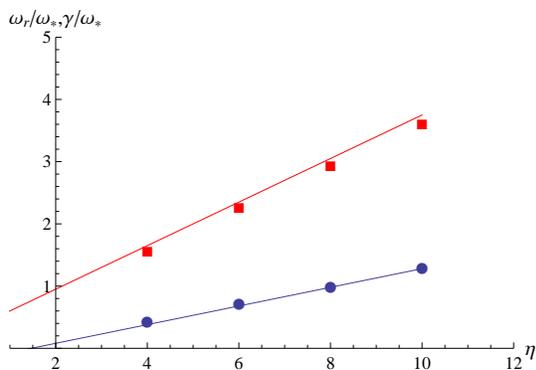}
\caption{Real (red squares) and imaginary (blue dots) part of the eigenfrequency {\it versus} $\eta = \xi$. Other parameters are $\nu = 10^{-2}, \beta = 10^{-2}, k_y = 0.05$. Straight lines are drawn for reference.}
\label{fig:dk2}
\end{figure}

We focus then on the role of plasma gradients, with no longer consideration of the drifts, $\fgb=0$; here the drift-kinetic formalism is expected to be more appropriate, and allows us to gain insight into the results of Fig.~\ref{fig:gammashear}. Taking simultaneously the limits $\eta,\,\xi\to\infty$ and $\nu\to 0$ considerably simplifies the equations above. Setting for further handiness $\eta=\xi$, in these limits the conductivity $\sigma$ shrinks within a layer of width $\Delta x\sim\rho_e$ around $x=0$, a physical consequence of the finite electron inertia. Being $\phi(0)=0$, the scalar potential is very small for $|x|\le\rho_e$, hence its correction to the r.h.s of Eq.~(\ref{eq:dka}) is negligible (at variance with the gyrokinetic approach, where $\phi$ plays a more important role). Eqns.~(\ref{eq:dka}) and (\ref{eq:dkp}) decouple, and we can fix $\omega$ by just solving the former. Inspection of this simplified equation shows that $\omega$ and $\eta$ do not enter individually but through their combination $\omega/\eta=\lambda$, which becomes now the proper eigenvalue. We can now compute $\lambda$, finding Im$(\lambda)>0$, which implies $\gamma$ to be linearly growing with $\eta$ for $\eta\gg 1$ (and the same dependence holds for $\omega_r$).

Quantitative results are shown in Fig.~\ref{fig:dk1}, where $\omega$ is shown as a function of the collisionality for fixed values of the other parameters, and $\eta=\xi \gg 1 $: a finite $\gamma>0$ is definitely recovered for $\nu\to 0$. Fig.~\ref{fig:dk2} shows the dependence of $\omega$ on $\eta=\xi$ for $\nu\ll 1$: the linear trend expected on the basis of the previous paragraph is found. Hence, also in this regime gyrokinetic results are recovered qualitatively. 


{\it Concluding remarks.} The linear stability of  MTMs represents a more multifaceted problem than previously thought; a unique ``collisionless regime'' cannot be singled out; rather, parameter regions with different stability properties can be identified depending on mechanisms such as particle inertia and non-ambipolar drifts. In particular the intrinsic high magnetic drifts of reversed field pinches and spherical tokamaks make such devices more prone to microtearing turbulence.

The authors wish to thank S. Cappello, X. Garbet and D. Dickinson for helpful hints and discussions, and the authors of GS2 for making the code publicly available. This work was supported by the European Communities under the contract of Association between EURATOM/ENEA.


\begin{thebibliography}{30}
\bibitem{hazeltine} R.~D. Hazeltine, D. Dobrott, and T.~S. Wang, Phys. Fluids {\bf 18}, 1778 (1975).
\bibitem{mast} D.~J. Applegate, C.~M. Roach, J.~W. Connor, S.~C. Cowley, W. Dorland, R.~J. Hastie, and N. Joiner, Plasma Phys. Control. Fusion \textbf{49}, 1113 (2007).
\bibitem{nstx} K.~L. Wong, S. Kaye, D. Mikkelsen, J. Krommes, K. Hill, R. Bell, and B. LeBlanc, Phys. Rev. Lett \textbf{99}, 135003 (2007). 
\bibitem{rfx} I. Predebon, F. Sattin, M. Veranda, D. Bonfiglio, and S. Cappello, Phys. Rev. Lett. \textbf{105}, 195001 (2010).
\bibitem{nonlin1} W. Guttenfelder, J. Candy, S.~M. Kaye, W.~M. Nevins, E. Wang, R.~E. Bell, G.~W. Hammett, B.~P. LeBlanc, D.~R. Mikkelsen, and H. Yuh, Phys. Rev. Lett. \textbf{106}, 155004 (2011).
\bibitem{nonlin2} H. Doerk, F. Jenko, M.~J. Pueschel, and D. Hatch, Phys. Rev. Lett. \textbf{106}, 155003 (2011). 
\bibitem{hmode} D. Dickinson, C.~M. Roach, S. Saarelma, R. Scannell, A. Kirk, and H.~R. Wilson, Phys. Rev. Lett. \textbf{108}, 135002 (2012)
\bibitem{drake} J.~F. Drake, N.~T. Gladd, C.~S. Liu, and C.~L. Chang, Phys. Rev. Lett. \textbf{44}, 994 (1980) .
\bibitem{gladd} N.~T. Gladd, J.~F. Drake, C.~L. Chang, and C.~S. Liu, Phys. Fluids \textbf{23}, 1182 (1980).
\bibitem{catto} P.~J. Catto and M.~N. Rosenbluth, Phys. Fluids \textbf{24}, 243 (1981)
\bibitem{garbet} X. Garbet, F. Mourgues and A. Samain, Plasma Phys. Control. Fusion \textbf{32}, 131 (1990).
\bibitem{vac2} I. Predebon and F. Sattin, 39th EPS Conference on Plasma Physics (Stockholm, 2012), \url{http://ocs.ciemat.es/epsicpp2012pap/pdf/P4.088.pdf}.
\bibitem{vac3} H. Doerk, F. Jenko, T. G\"orler, D. Told, M.~J. Pueschel, and D.~R. Hatch, Phys. Plasmas \textbf{19}, 055907 (2012).
\bibitem{vac4} D. Dickinson, C.~M. Roach, S. Saarelma, R. Scannell, A. Kirk, H.~R. Wilson, \url{http://arxiv.org/abs/1209.3695}.
\bibitem{vac5} D. Carmody, M.~J. Pueschel, P.~W. Terry, \url{http://arxiv.org/abs/1301.4576}.
\bibitem{vac6} D.~R. Hatch, M.~J. Pueschel, F. Jenko, W.~M. Nevins, P.~W. Terry, and H. Doerk, Phys. Rev. Lett. \textbf{108}, 235002 (2012); D.~R. Hatch, M.~J. Pueschel, F. Jenko, W.~M. Nevins, P.~W. Terry, and H. Doerk, Phys. Plasmas \textbf{20}, 012307 (2013).
\bibitem{gs2} M. Kotschenreuther, G. Rewoldt, and W. M. Tang, Comput. Phys. Commun. \textbf{88}, 128 (1995);
  W. Dorland, F. Jenko, M. Kotschenreuther, and B.~N. Rogers, Phys. Rev. Lett. \textbf{85}, 5579 (2000).
\bibitem{gs2rfp} I. Predebon, C. Angioni, and S.~C. Guo, Phys. Plasmas \textbf{17}, 012304 (2010).
\bibitem{hall} K. Hallatschek and W. Dorland, Phys. Rev. Lett. \textbf{95}, 055002 (2005).
\bibitem{gutten}  W. Guttenfelder, J. Candy, S.~M. Kaye, W.~M. Nevins, R.~E. Bell, G.~W. Hammett, B.~P. LeBlanc, and H. Yuh, Phys. Plasmas \textbf{19}, 022506 (2012).
\end{thebibliography}
\end{document}